\documentclass[a4paper]{article}
\usepackage{graphicx}
\usepackage{indentfirst}
\usepackage{booktabs}
\usepackage{authblk}
\usepackage{latexsym}
\usepackage{amsmath}
\usepackage{mathrsfs}
\usepackage{hyperref}
\usepackage{wrapfig}

\makeatletter
\renewcommand{\fnum@figure}{Fig. \thefigure.\@gobble}
\makeatother

\title{Size-Dependent Tensile Behavior and Dislocation Dynamics in Cu and Ag Nanowires: A Molecular Dynamics Study}
\author[1,2]{Xiaorui Hu\thanks{Corresponding author: 20223984@stu.neu.edu.cn}}
\author[1,2]{Jiawei Xiong}
\affil[1]{School of Materials Science and Engineering, Northeastern University, Shenyang 110819, China}
\affil[2]{Key Laboratory for Anisotropy and Texture of Materials, Ministry of Education, Northeastern University, Shenyang 110819, China}
\begin{document}

\maketitle

\begin{abstract}
By using molecular dynamics simulations, the research examine how copper and silver nanowires respond to tensile loading in order to clarify their nanoscale deformation mechanisms. The results demonstrate that these two metal nanowires follow notably different stress–strain trends, with silver wires exhibiting greater elastic stiffness and higher yield points at equivalent diameters—an effect likely rooted in silver’s stronger atomic bonding and more stable microstructure. A pronounced size effect is observed: as the wire diameter diminishes, both the yield strength and ultimate tensile strength increase substantially, a behavior driven by the higher proportion of surface atoms that enhance dislocation nucleation and mobility. Atomistic analyses further underscore the dominant role of dislocations during plastic deformation, and in particular, reveal that surface‐initiated dislocations in thinner wires critically affect their fracture behavior. 
\end{abstract}

\section{Introduction}
Metal nanowires have captured lots of attention due to their potential applications in microelectronics, optoelectronics, catalysis, and other disciplines. Copper and silver nanowires, in particular, have emerged as focal points of research, courtesy of their outstanding electrical and thermal conductivities. Nevertheless, the mechanical properties of materials at the nanoscale diverge remarkably from those of macroscopic counterparts, thereby necessitating an in-depth comprehension of their mechanical behaviors during the tensile process.

Lu et al. \cite{lu2011molecular} conducted molecular dynamics simulations to probe into the tensile mechanical properties of single-crystal copper nanowires with distinct crystal orientations (<100>, <110>, <111>). Their findings revealed that copper nanowires oriented along the <111> direction exhibited the highest yield strength, whereas those with the <100> orientation boasted the largest yield strain. This clearly demonstrates that crystal orientation exerts a substantial impact on the plastic deformation behavior of copper nanowires.

Wang et al. \cite{xizhi2024study} employed molecular dynamics simulations to scrutinize the torsional deformation behaviors of copper nanowires under varying crystal orientations, different ratios of inner to outer diameters within the crystal lattice, and diverse numbers of twin interfaces. The results indicated that diminishing the ratio of inner to outer diameters and reducing the number of twin interfaces could enhance the torsional mechanical properties of copper nanowires. Yuan et al. \cite{yuan2014molecular} simulated the tensile deformation behaviors of polycrystalline silver nanowires with different grain sizes via the molecular dynamics approach. It was ascertained that when the grain size was less than 13.49 nm, the polycrystalline silver nanowires underwent a softening phenomenon, presenting an inverse Hall-Petch relationship. Here, the plastic deformation was predominantly governed by grain boundary sliding and grain rotation. In contrast, when the grain size exceeded 13.49 nm, dislocation slip became the principal mode of plastic deformation, and copious twin structures emerged in the later stages of deformation. These discoveries have proffered novel insights into understanding the mechanical properties of silver nanowires. Evidently, the mechanical properties of copper and silver metal nanowires are modulated by factors including crystal orientation, grain size, and defects. Molecular dynamics simulations enable a profound understanding of how these factors shape the plastic deformation behaviors of nanowires.
\section{Computational algorithm}
In the academic exploration of the mechanical properties of metal nanowires, molecular dynamics (MD) simulation, as a pivotal computational methodology, 
is capable of providing high-precision atomic-scale information, thereby uncovering the microscopic mechanical behaviors of materials. Through MD simulations, critical mechanical characteristics, such as the tensile, fracture, and stress-strain relationships of metal nanowires under external forces, can be meticulously investigated. Notably, at the nanoscale, metal nanowires exhibit mechanical behaviors that deviate from those of macroscopic materials, primarily due to the pronounced impacts of surface and size effects.

During MD simulations of metal nanowire tension, the construction of an apt metal nanowire model is prerequisite. In this research, periodic boundary conditions are harnessed to simulate infinitely long nanowires. To accurately characterize the interactions between metal atoms, potential energy functions are routinely utilized. Specifically, the embedded atom method (EAM) potential \cite{williams2006embedded} is adopted herein to simulate the interatomic interactions of metals. The total
potential energy of the system $E_total$ is determined by
\begin{equation}
	\begin{aligned}
		& E_{\text {total }}=\frac{1}{2} \sum_{i j} V_{i j}\left(r_{i j}\right)+\sum_i F_i\left(\bar{\rho}_i\right) \\
		& \bar{\rho}_i=\sum_{j \neq i} \rho_j\left(r_{i j}\right)
	\end{aligned}
\end{equation}
where $V_{i j}\left(r_{i j}\right)$ is the potential energy between atoms $i$ and $j$ having a distance of $r_{i j}, F_i\left(\overline{\rho_i}\right)$ is the embedded energy with an electron density of $\bar{\rho}_i$ at the position of the atom $i$. The density value is obtained from the superposition and sum of the electron density from the nearest neighboring atoms of the atom $i . \rho_j$ is the electron density of the atom $j$.

The tensile process is effected at both ends of the nanowire by applying a uniform external force or strain rate, with the simulated temperature and pressure conditions tailored to the specific research requirements. To delve into diverse mechanical properties, the researchers focus on indices such as the stress-strain curves, elastic modulus, yield strength, and fracture behavior of the nanowires.

Throughout the tensile process, the deformation modes of metal nanowires typically encompass three stages: elastic deformation, plastic deformation, and fracture. It has been empirically observed that as the size of metal nanowires decreases, both their elastic modulus and yield strength manifest an evident upward trend, which contrasts with the common trends witnessed in macroscopic materials. This phenomenon can be chiefly ascribed to the substantial specific surface area effect and the contribution of surface energy borne by the surface atoms of the nanowires. Additionally, metal nanowires frequently exhibit relatively intricate plastic deformation behaviors during the tensile process, including the generation, propagation, and interaction of dislocations. Particularly at smaller scales, metal nanowires are prone to the activation of surface dislocations, which subsequently impacts their fracture behaviors. A wealth of studies has attested that the fracture of nanowires often transpires not through classical dislocation slip or fracture modalities but via the formation and propagation of surface cracks.

This investigation endeavors to elucidate the mechanical characteristics of Cu and Ag nanowires subjected to various tensile conditions through computational molecular dynamics (MD) approaches.  The researchers particularly focus on examining the mechanical responses, deformation pathways, and nanoscale property variations between these materials.  Throughout their examination, they systematically manipulate variables including strain rates, thermal conditions, and dimensional parameters to conduct comprehensive simulations of nanowire stretching processes, thereby revealing atomic-level reactions at microscopic scales.  The fundamental research encompasses: an extensive comparative examination of Cu and Ag nanowires' mechanical behaviors during elongation, analyzing stress-strain relationships across different strain rates and thermal environments, followed by thorough assessment of tensile strength, elasticity moduli, and fracture properties of both materials;  an exploration of dimensional influences on the mechanical attributes of these nanowires by simulating structures with varying diameters to understand how size variations affect material properties such as yield strength and fracture resistance;  and a detailed exposition of the microscopic deformation mechanisms by observing atomic-scale transformations to uncover the underlying processes during tensile deformation, including dislocation movement and twinning phenomena.  The investigators employ molecular dynamics simulation methodologies to systematically analyze the mechanical properties of copper nanowires (Cu NWs) and silver nanowires (Ag NWs) during tensile deformation.  Their methodological approach involves constructing atomic models based on face-centered cubic crystal structures, with nanowire diameters specified at 3.5 nm, 4 nm, and 4.5 nm, while lengths are determined according to experimental requirements ensuring sufficient extension space.  Axial periodic boundary conditions are implemented to simulate infinitely extended nanowires, whereas non-periodic conditions are applied radially to eliminate unwanted interactions between neighboring structures. The detailed implementation is elaborated as follows:

Firstly, the cluster radius is given as 
\begin{equation}
	r_0=\frac{n\cdot a_{0}}{2}
\end{equation}
 Then, the lattice basis vector matrix is constructed as 
\begin{equation}
	b_0(i, j)= \begin{cases}n \cdot a_0, & i=j \\ 0 \quad, & i \neq j
	\end{cases}
\end{equation}
Finally, the relative coordinates are converted into actual ones via the lattice basis vectors.
\begin{equation}
	x_1=x_0\cdot b_0(1,1)
\end{equation}
The initial structure is energy-minimized to ensure that the system is in a stable state, and an appropriate number of grain boundaries and dislocations are introduced to simulate the defects in real materials.
To accurately describe the interactions between copper and silver atoms, the Embedded Atom Method (EAM) potential energy function is selected in this study. The EAM potential can effectively capture the bonding characteristics between metals and is suitable for describing the mechanical behaviors of metal nanowires with FCC structures.Specifically, the EAM potential parameterized by Mendelev et al. is adopted to ensure the reliability and accuracy of the simulation results.

During the simulation, the time step is set to 1 fs to balance computational efficiency and accuracy. The system temperature is controlled within the range of 300 K to 350 K by the Nose-Hoover thermostat to simulate the influence of different temperatures on the mechanical properties of nanowires. The tensile process adopts the constant strain rate method, and nanowires with different diameters are generated to study the influence of the size effect on the mechanical properties of nanowires.The key equations of the Nose-Hoover thermostat are as follows:
\begin{equation}
	\begin{aligned}
		&\dot{r}_i = \frac{\mathbf{p}_i}{m_i} \\
		&\dot{\mathbf{p}}_i = \mathbf{F}_i - \frac{p_\eta}{Q} \mathbf{p}_i \\
		&\dot{\eta} = \frac{p_\eta}{Q} \\
		&\dot{p}_\eta = \sum_{i=1}^N \frac{\mathbf{p}_i^2}{m_i} - dNk_BT
	\end{aligned}
\end{equation}

The Nose-Hoover thermostat controls the temperature by introducing an additional degree of freedom, $\eta$ , and its conjugate momentum, $p_\eta$ . The term$ - \frac{p_\eta}{Q}\mathbf{p_i} $in the momentum update equation acts as a frictional force that adjusts the velocities of the particles. When the kinetic energy of the system is higher than the target value corresponding to the desired temperature, this term slows down the particles, reducing the kinetic energy and thus the temperature. Conversely, when the kinetic energy is lower than the target value, the frictional force helps to increase the velocities, raising the temperature. Over time, this mechanism allows the system to reach and maintain a temperature close to the target temperature T . The parameter $Q$ determines the relaxation rate of the thermostat dynamics, influencing how quickly the system approaches the target temperature.

During the tensile process, an external force is applied along the X-axis of the nanowire, and the strain control method is used to achieve gradual stretching. Specifically, one end of the nanowire is fixed, and the other end is moved at a constant rate to simulate the stress-strain response under actual tensile conditions. To reduce the influence of boundary effects on the simulation results, constraints are applied or free expansion is allowed in the radial direction to enable the natural deformation of the nanowire in non-axial directions.

 The atomic positions, velocities, stress-strain data, etc. of the system are recorded in real time in the progress. Through post-processing analysis, the stress-strain curve is used to calculate key mechanical parameters such as the elastic modulus, yield strength, and fracture strength. In addition, Failure Mode Analysis is utilized to observe microscopic deformation mechanisms such as atomic rearrangement, dislocation formation and movement, and twin transformation during the tensile process of nanowires. Visualization tools are used to display the atomic-level deformation process intuitively, so as to gain an in-depth understanding of the internal mechanisms of mechanical properties.

To comprehensively evaluate the mechanical properties of nanowires, simulations are carried out under different size and temperature conditions. By comparing the mechanical responses of nanowires with different diameters under the same tensile conditions, the influence law of size on mechanical properties is revealed. At the same time, the stress-strain behaviors of nanowires at different temperatures are analyzed to explore the sensitivity of material mechanical properties to temperature changes and the underlying physical mechanisms.
\newpage
\section{Results and Discussion}
\subsection{An Overview of Tensile Progress  }
A tensile force is applied in the [100] direction, and the tensile process is recorded to obtain the stress-strain diagrams of the metal nanowires.
\begin{figure}[h]
	\includegraphics[width=\columnwidth]{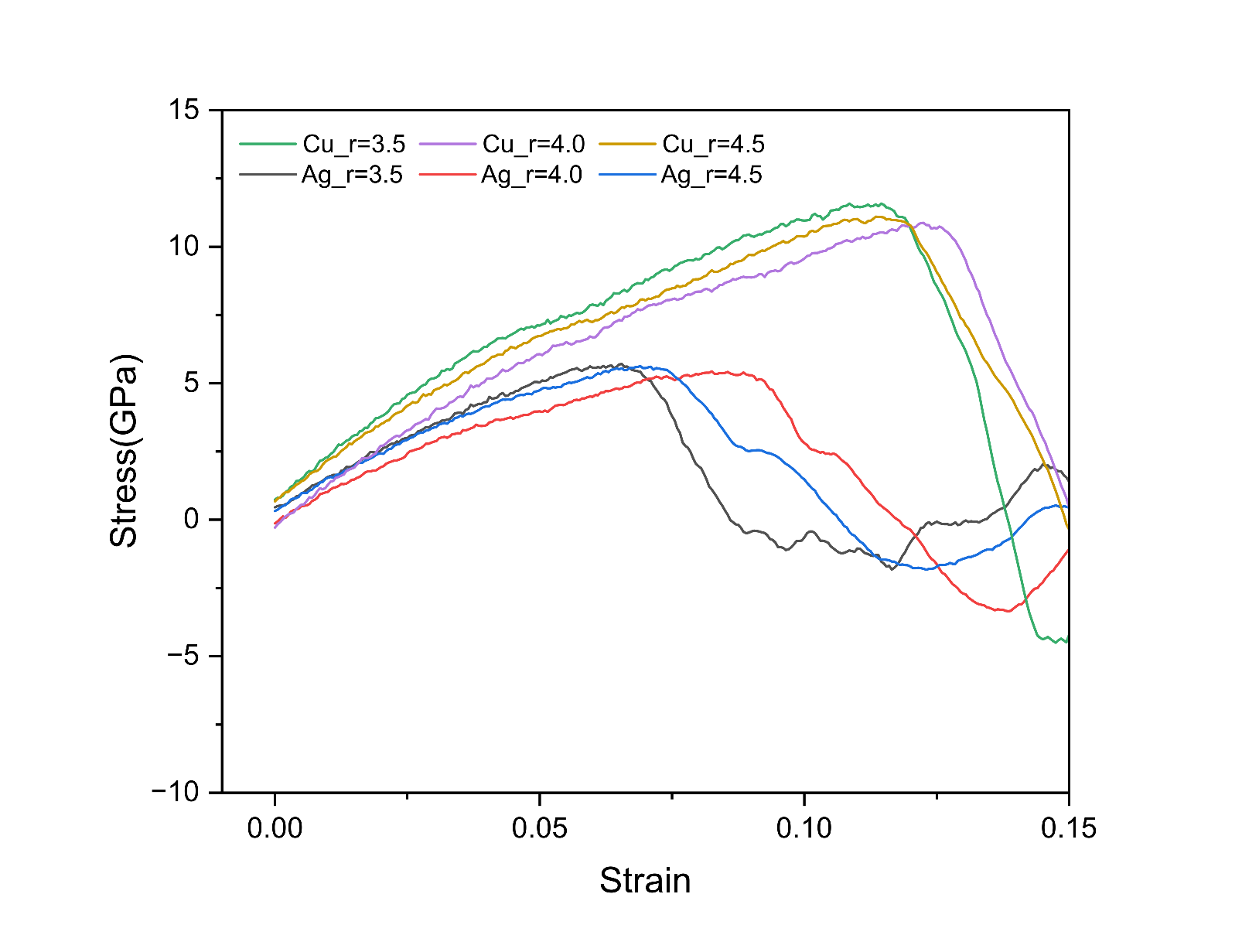}
	\caption{Stress-strain diagrams of silver and copper nanowires with different sizes under a load at T = 300 K.}
	\label{pic1}
\end{figure}

According to Figure \ref{pic1}, in the initial part of the stress-strain curve, all nanowires exhibit a linear stress-strain relationship. This stage conforms to Hooke's law:
\begin{equation}
	\sigma= E \cdot \epsilon
\end{equation}
where $\sigma$ is stress, $\epsilon$ is strain, and $E$ is the elastic modulus of the material.
 
At this stage, the atomic spacing inside the material undergoes a slight change, and the bond lengths between atoms are stretched or compressed, but the atoms do not have permanent displacements. When the external force is removed, the material can fully recover to its original shape. As the strain increases, the stress-strain curve begins to deviate from the linear relationship, and the yield point appears. The yield point marks the transition of the material from elastic deformation to plastic deformation. Near the yield point, dislocations start to be generated inside the material, and the generation and movement of dislocations are the main mechanisms of plastic deformation.
After entering the plastic deformation stage, the stress-strain curve is no longer linear. The dislocations inside the material start to proliferate and move in large numbers, causing irreversible deformation of the material. When the stress reaches its maximum value (tensile strength), the material begins to show the necking phenomenon. Necking refers to the reduction of the cross-sectional area in local regions of the material.
 
As the necking develops, the stress-strain curve begins to decline, eventually leading to the fracture of the material.
It can be seen from the figure that the slopes (i.e., elastic moduli) of silver nanowires with different diameters are different in the elastic deformation stage. Generally speaking, silver nanowires with larger diameters have higher elastic moduli. This is because larger-diameter nanowires have more atoms and stronger interatomic bonding, which can more effectively resist elastic deformation.

As the diameter of the silver nanowire increases, both its yield strength and tensile strength increase significantly. For example, when comparing silver nanowires of [specific diameters], the nanowire with a larger diameter can withstand higher stress before it starts to yield and fracture. This increase in strength can be attributed to the more stable atomic structure and stronger interatomic bonding of larger-diameter nanowires, which can more effectively resist external forces.

In the plastic deformation stage, the strain hardening phenomenon of silver nanowires with larger diameters is more obvious. This is because the proliferation and interaction of dislocations inside larger-diameter nanowires are more complex, causing the strength of the material to increase rapidly with the increase of plastic deformation.

Similar to silver nanowires, the elastic modulus of copper nanowires also increases with the increase of diameter. The linear relationships of copper nanowires with different diameters in the elastic deformation stage indicate that their elastic deformation mechanisms are mainly determined by interatomic bonding. The yield strength and tensile strength of copper nanowires also increase with the increase of diameter. However, at the same diameter, the yield strength and tensile strength of copper nanowires are generally lower than those of silver nanowires. This is due to the differences in atomic structures and interatomic bonding strengths between silver and copper. Silver has stronger interatomic bonding, enabling it to withstand higher stress during the tensile process.

Copper nanowires also exhibit the strain hardening phenomenon [5] in the plastic deformation stage. However, compared with silver nanowires, the strain hardening rate of copper nanowires may be different, depending on their atomic structures and dislocation movement mechanisms.
\subsection{Strength - Size Relationship}
Both silver and copper nanowires exhibit a tendency for their strength to increase with the growth of diameter. This phenomenon can be explained from the following aspects:

\textbf{Atomic Quantity and Bonding:} Nanowires with larger diameters contain more atoms, and the bonding between these atoms is more numerous and stronger, enabling them to resist external forces more effectively.

\textbf{Dislocation Movement Restriction:} In larger-diameter nanowires, the movement of dislocations is hindered by more atoms, making it more difficult for dislocations to proliferate and move, thus enhancing the material's strength.

Generally speaking, nanowires with smaller diameters are more prone to plastic deformation. This is because the interatomic bonding in small-diameter nanowires is relatively weak, and the proportion of surface atoms is relatively high. Surface atoms possess higher energy and are more likely to diffuse and react, which impacts the mechanical properties of nanowires and renders them more susceptible to plastic deformation.
\subsection{Microstructure and Dislocation}
After the elastic deformation stage ends, dislocations start to be generated inside the material. Dislocations are local irregularities in the atomic arrangement within a crystal, and their movement is the primary mechanism for plastic deformation.

For nanowires, due to their small size, the generation and movement of dislocations are restricted. The surface and interface effects of nanowires can influence the generation and propagation of dislocations. Nanowires, because of their small size, have a relatively high proportion of surface atoms. These surface atoms have high energy and are more likely to diffuse and react. The presence of surface atoms affects the mechanical properties of nanowires, making them more likely to undergo plastic deformation and reducing their strength and hardness. At the nanoscale, the mechanical properties of materials differ significantly from those of macroscopic materials. The size effect may lead to significant changes in the strength and plasticity of nanowires, and sometimes even result in opposite outcomes.
\begin{figure}[h]
	\centering
	\includegraphics[width=\columnwidth]{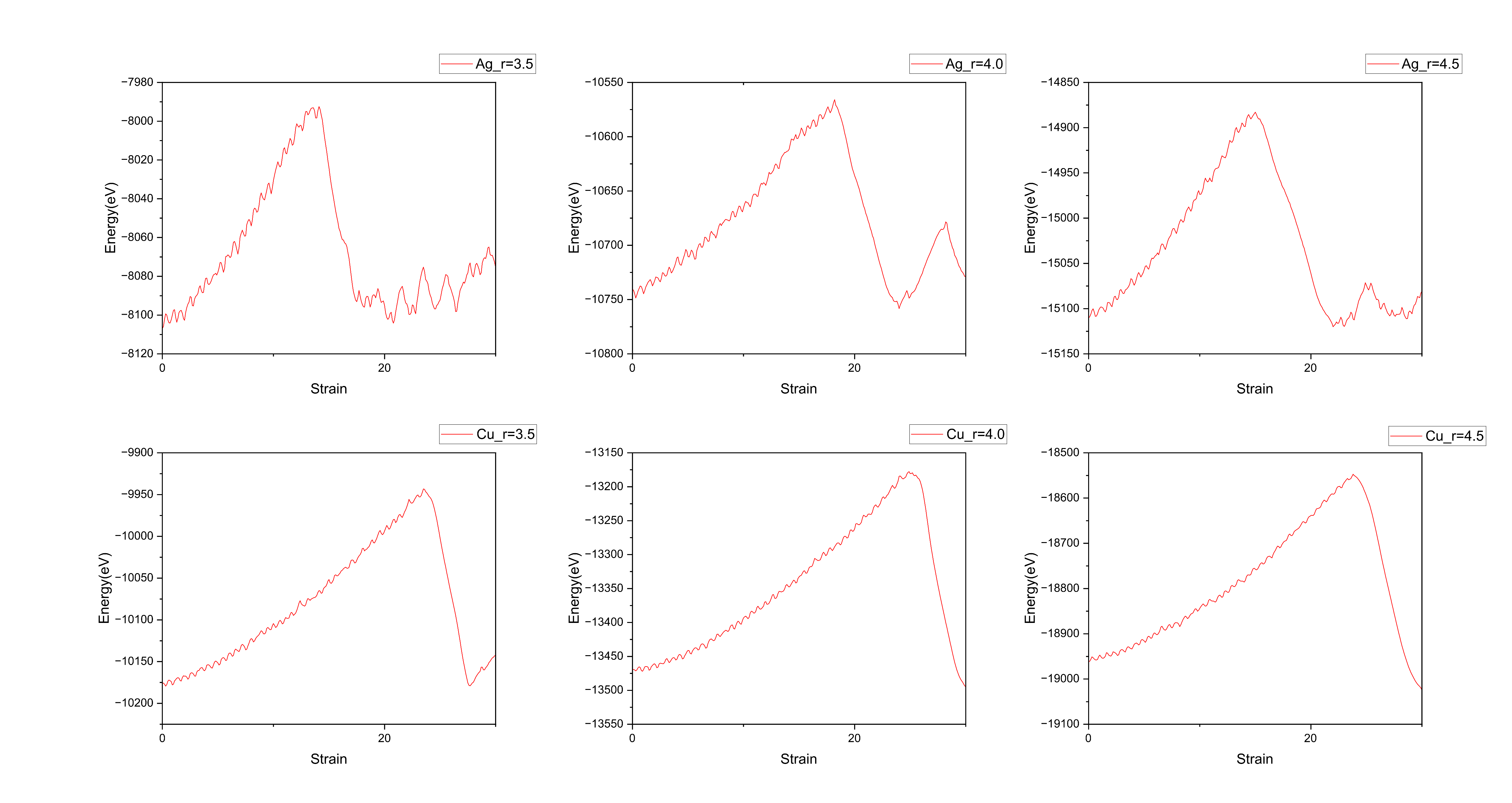}
	\caption{Energy-strain diagrams of silver and copper nanowires of different sizes, loading at T = 300K}
	\label{pic2}
\end{figure}

During the elastic deformation stage (which can be seen as the linear stage from the stress-strain diagram), the energy-strain curve should take the form of a quadratic function. This is because the elastic strain energy is proportional to the square of the strain. As the strain increases, the elastic strain energy gradually accumulates.

When the material enters the plastic deformation stage (where the stress-strain curve deviates from linearity), the slope of the energy-strain curve begins to change. In the plastic deformation stage, the increase in energy comes not only from further accumulation of elastic strain energy but also from the energy consumed during the plastic deformation process. At this time, the slope of the energy-strain curve will be greater than that in the elastic stage.

As can be seen from Figure \ref{pic1}, for silver nanowires under the same strain, the stress value is higher as the diameter increases. This means that in the energy-strain diagram, silver nanowires with larger diameters absorb more energy under the same strain. During the elastic deformation stage, due to the elastic modulus, the elastic modulus of silver nanowires determines the rate of elastic strain energy accumulation. Silver nanowires with larger diameters have a higher elastic modulus, so they accumulate elastic strain energy more quickly. In the plastic deformation stage, silver nanowires with larger diameters can withstand greater stress due to their higher strength, consuming more energy during the plastic deformation process. This will result in a more significant increase in energy for larger-diameter silver nanowires during the plastic deformation stage in Figure \ref{pic2}.

Similar to silver nanowires, an increase in the diameter of copper nanowires also leads to an increase in the absorbed energy under the same strain. However, under the same diameter, the stress-strain curve of copper nanowires is lower than that of silver nanowires, which means that copper nanowires absorb less energy than silver nanowires under the same strain. In Figure \ref{pic2}, both the accumulation of elastic strain energy and the consumption of plastic deformation energy in copper nanowires are less than those in silver nanowires under the same strain.

For nanowires of the same metal, as the diameter increases, during the elastic deformation stage, due to the increase in the elastic modulus, the elastic strain energy accumulates more quickly, and the slope of the energy-strain curve is larger. In the plastic deformation stage, due to the increase in strength, more plastic deformation energy is consumed, and the energy-strain curve rises more steeply during plastic deformation. Nanowires with smaller diameters are more prone to plastic deformation, which means that in Figure 2, nanowires with smaller diameters will show a significant change in the slope of the energy-strain curve (transition from the elastic stage to the plastic stage) at a lower strain.

During the elastic deformation stage, energy is mainly stored in the elastic deformation of the interatomic bonds. When the nanowire is stretched, the atomic spacing changes, and the interatomic bonds are elongated or compressed, storing elastic strain energy. After entering the plastic deformation stage, dislocations start to move. The movement of dislocations needs to overcome energy barriers such as lattice resistance, consuming energy. Therefore, in Figure \ref{pic2}, the change in the rate of energy increase during the plastic deformation stage reflects the difficulty of dislocation movement and the energy consumption situation.

For nanowires with different diameters, the internal atomic arrangement of larger-diameter nanowires is more regular, and dislocation movement needs to overcome more interatomic interactions, resulting in more energy consumption during the plastic deformation process. The small size of nanowires leads to a relatively high proportion of surface atoms. Surface atoms have high energy, and during the stretching process, the movement and deformation of surface atoms consume energy. The size effect makes the mechanical properties of nanowires different from those of macroscopic materials. In Figure \ref{pic2}, due to the surface effect and size effect, nanowires with smaller diameters may show non-linear changes in the energy-strain curve at a lower strain, which reflects their earlier entry into the plastic deformation stage and different energy absorption mechanisms. 
\begin{figure}
	\centering
	\includegraphics[width=0.8\columnwidth]{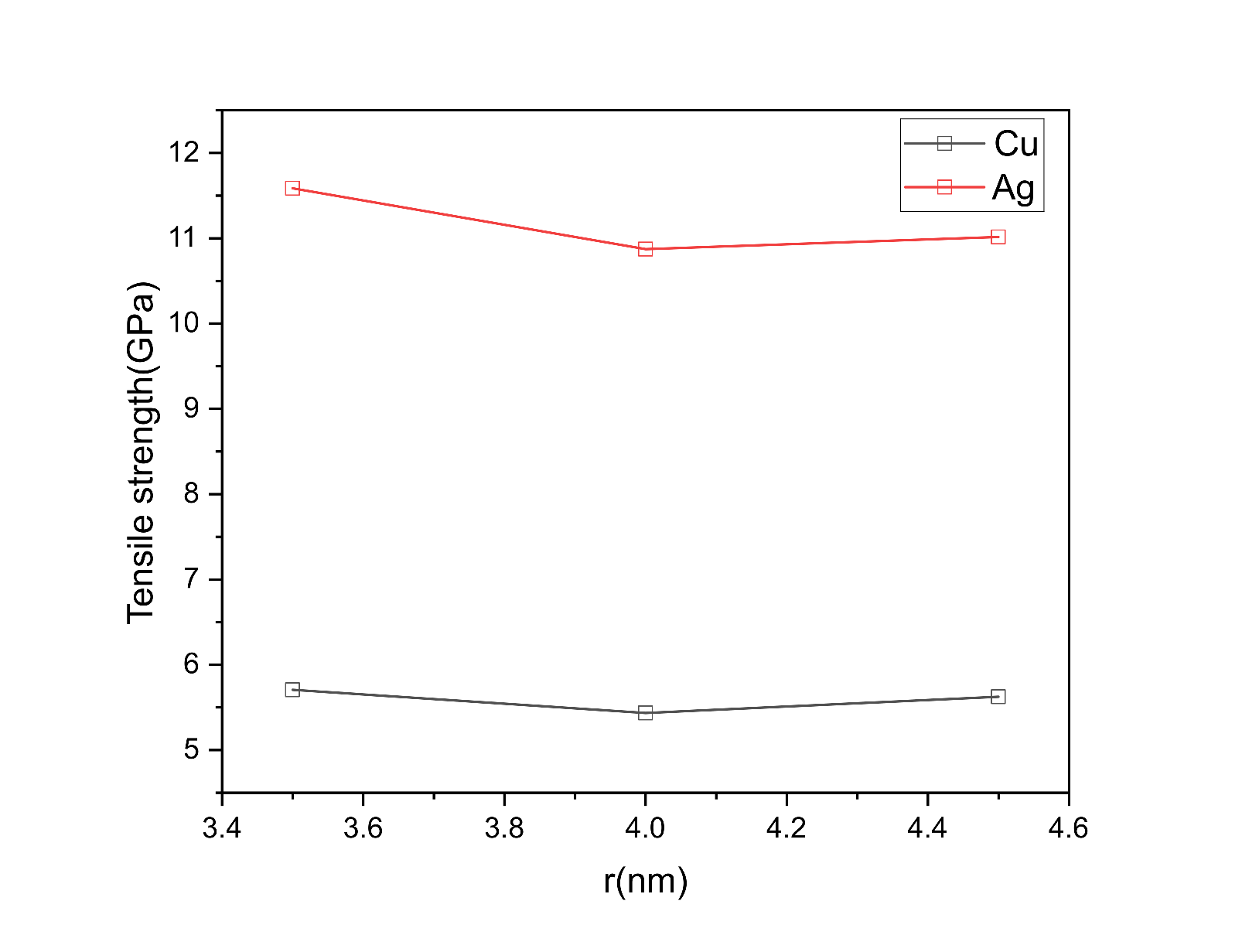}
	\caption{Tensile strength of silver and copper nanowires of different sizes}
	\label{pic3}
\end{figure}

It can be seen from Figure \ref{pic3} that the tensile strengths of both silver (Ag) and copper (Cu) nanowires increase with the growth of diameter. Throughout the entire diameter range (3.4 nm - 4.6 nm), the tensile strength of silver nanowires is consistently higher than that of copper nanowires.

When the diameter increases from 3.4 nm to 4.6 nm, the tensile strength of silver nanowires drops slightly from approximately 11.5 GPa to around 11 GPa. Despite this slight downward trend, the tensile strength generally remains at a relatively high level. Such high tensile strength may be attributed to the bonding strength between silver atoms and the microstructural characteristics of silver nanowires. The bonding between silver atoms can effectively resist external forces during the stretching process, enabling the nanowires to maintain high strength even at larger diameters.

The tensile strength of copper nanowires also shows an upward trend as the diameter increases. It rises from about 5.5 GPa at a diameter of 3.4 nm to approximately 6 GPa at a diameter of 4.6 nm. Compared with silver nanowires, the tensile strength of copper nanowires is lower. This may be due to the relatively weaker bonding strength between copper atoms or the fact that the microstructure of copper nanowires is more prone to deformation during the stretching process.

For both types of metal nanowires, the increase in diameter leads to an increase in tensile strength. This phenomenon can be explained from the following two aspects:
Atomic Quantity and Bonding: As the diameter of the nanowire increases, the number of atoms rises, and the number of bonds between atoms also increases accordingly. More bonds can resist external forces more effectively, thus enhancing the tensile strength of the nanowire.

Microstructural Stability: Nanowires with larger diameters usually possess more stable microstructures [7], which helps maintain the integrity of the material during the stretching process and further improves the tensile strength.
Combined with the previous stress-strain diagrams, it can be found that Figure \ref{pic3} corresponds to the maximum stress (tensile strength) in Figure \ref{pic1}. In Figure \ref{pic1}, nanowires with larger diameters can withstand higher stress before starting to yield and fracture, which is consistent with the trend in this figure that the larger the diameter, the higher the tensile strength.
\newpage
\begin{figure}
	\centering
	\includegraphics[width=0.7\columnwidth]{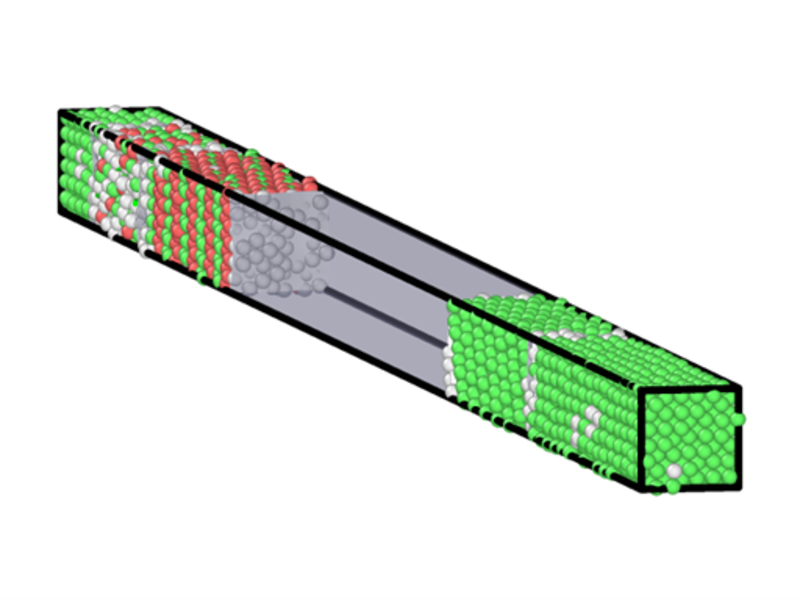}
	\caption{Schematic diagram of dislocation analysis after the fracture of copper nanowires}
	\label{pic4}
\end{figure}

Figure \ref{pic4} shows the nanowire after being stretched to fracture. During the tensile process of nanowires, the generation and movement of dislocations are important factors influencing the tensile strength [6]. For nanowires with smaller diameters, dislocations are more likely to be generated and move on the surface or inside because the smaller size limits the dislocation-blocking mechanisms. As the diameter increases, the generation and movement of dislocations are hindered by more atoms, enabling the nanowires to withstand higher stress and thus increasing the tensile strength.
\section{Conclusions}
In this study, through molecular dynamics simulations, the researchers have thoroughly explored the mechanical properties of copper and silver nanowires under different tensile conditions, uncovering their unique behaviors and deformation mechanisms at the nanoscale. The research findings indicate that the mechanical properties of metal nanowires are influenced by multiple factors, including the size effect, temperature variation, and the atomic structural characteristics of the materials themselves.

Firstly, it was found that there are significant differences in the stress-strain responses of copper and silver nanowires during the tensile process. Silver nanowires exhibit higher elastic moduli and yield strengths at the same diameter, which is closely related to the stronger bonding forces between silver atoms and their more stable microstructures. As the diameter of the nanowire increases, the tensile strength of silver nanowires remains at a relatively high level, while that of copper nanowires is relatively low, demonstrating the essential differences in the micromechanical behaviors of the two materials.

Secondly, the size effect plays a crucial role in the mechanical properties of metal nanowires \cite{wu2013nanostructure}. Research shows that as the diameter of the nanowire decreases, the yield strength and tensile strength of the material increase significantly. This phenomenon can be attributed to the increased proportion of surface atoms in nanowires with smaller diameters. These surface atoms have higher energy and are more likely to generate and move dislocations, thus affecting the plastic deformation behavior of the material.

In addition, this paper also explored the microscopic deformation mechanisms of nanowires. By observing the atomic-level deformation processes, it was discovered that the generation, propagation, and interaction of dislocations play important roles in the plastic deformation stage. Especially at smaller scales, the excitation of surface dislocations significantly affects the fracture behavior of nanowires \cite{liang2005shape}, resulting in a fracture mode that is remarkably different from that of macroscopic materials.

To sum up, this study not only provides important theoretical bases for understanding the mechanical properties of metal nanowires but also lays the foundation for future research on the design and application of nanomaterials. Through further experimental and simulation studies, it is expected that the mechanical behaviors of nanomaterials under different conditions can be more comprehensively revealed, providing guidance for the development of new nanomaterials.

\bibliographystyle{plain}
\bibliography{reference.bib}
\end{document}